# Unsupervised Classification of Variable Stars

Lucas Valenzuela[1]★
Karim Pichara[1,2]†

[1]*School of Engineering, Computer Science Department, Pontificia Universidad Católica de Chile*
[2] *Millenium Institute of Astrophysics*

**ABSTRACT**

During the last ten years, a considerable amount of effort has been made to develop algorithms for automatic classification of variable stars. That has been primarily achieved by applying machine learning methods to photometric datasets where objects are represented as light curves. Classifiers require training sets to learn the underlying patterns that allow the separation among classes. Unfortunately, building training sets is an expensive process that demands a lot of human efforts. Every time data comes from new surveys; the only available training instances are the ones that have a cross-match with previously labelled objects, consequently generating insufficient training sets compared with the large amounts of unlabelled sources. In this work, we present an algorithm that performs unsupervised classification of variable stars, relying only on the similarity among light curves. We tackle the unsupervised classification problem by proposing an untraditional approach. Instead of trying to match classes of stars with clusters found by a clustering algorithm, we propose a query based method where astronomers can find groups of variable stars ranked by similarity. We also develop a fast similarity function specific for light curves, based on a novel data structure that allows scaling the search over the entire dataset of unlabelled objects. Experiments show that our unsupervised model achieves high accuracy in the classification of different types of variable stars and that the proposed algorithm scales up to massive amounts of light curves.

## 1 INTRODUCTION

In the last decade, lots of efforts have been made in automating the classification of variable stars (Debosscher et al. 2007; Richards et al. 2011; Kim et al. 2011; Bloom et al. 2012; Pichara et al. 2012; Pichara & Protopapas 2013; Kim et al. 2014; Nun et al. 2014, 2015; Mackenzie et al. 2016; Pichara et al. 2016; Cabrera-Vives et al. 2017; Elorrieta et al. 2016; Huijse et al. 2014; Ivezic et al. 2008; Förster et al. 2016). That has been widely achieved by applying machine learning methods to photometric datasets where objects are represented by their light curves. Machine learning classifiers have been used for this purpose. Classifiers can automatically identify classes of unlabelled data after learning patterns from datasets where the classes—or labels—are already known. This learning stage is called *training* and the initially labelled dataset, from which the classifier learns, is consequently called *training set*. Unfortunately, building training sets for light curves is a costly process that demands a lot of human efforts (Bhamidipaty et al. 2002). According to Angeloni et al. (2014), light curves have to meet the following conditions to be considered good enough for training: (1) a relative error of less than 1/10 of the light curve amplitude and (2) no significant gaps in the phase coverage. Also, all templates must be in the same band as the light curves in the dataset to be classified.

Angeloni et al. (2014) present the following alternatives to obtain training instances: (1) searching on previous surveys for labelled light curves that meet these conditions, (2) collecting labels by cross-matching a labelled dataset with one that contains light curves meeting the other requirements, (3) setting up an observational program to build light curves in the correspondent band, targeting previously studied stars. After obtaining labelled light curves by either of these methods, collected objects have to be carefully examined by an astronomer to determine if they meet the expected quality.

In most of the cases, when data come from new surveys, the only available training instances are the ones obtained by cross-matching with previously labelled objects. As a consequence, the acquired training sets are weak compared to the large amounts of unlabelled sources (Richards et al. 2011). In addition, in most of the cases astronomers are interested in the identification of a particular class of variable stars, and for them, it is pointless to wait for a training set that contains labelled data from other classes of variability.

★ Contact e-mail: luvalenz@uc.cl
† Contact e-mail: kpb@ing.puc.cl

Our work offers an algorithm able to perform a fast identification of variable stars of any class of interest, relying only on the similarity among objects. Typical Machine Learning approaches that perform unsupervised classification are based on clustering algorithms (T. et al. 1996; Z. 1998; A. 1999; Bradley & Mangasarian 1998) that try to match clusters with classes.

Commonly, clustering algorithms used to find classes perform several iterations over the data, also need to tune up several parameters and define a suitable distance/similarity function that in many cases is very hard to find. Unfortunately, those drawbacks make very hard to apply clustering techniques to variable stars classification directly. In this work, we address the unsupervised classification problem by proposing a different approach. First, instead of performing several iterations over the data to discover classes using clusters, we propose a query based method where astronomers can find groups of variable stars ranked by similarity. Second, we develop a fast distance function specific for light curves, based on a novel data structure that allows scaling the search over the entire dataset of unlabelled objects.

A classical unsupervised approach would first run a clustering algorithm over the data and then would attempt to identify the classes of stars, by manually inspecting the objects within the clusters. In general, the resulting clusters fail to capture objects from just one class, making hard to separate the instances inside every cluster. To overcome that problem, our query based approach allows astronomers to provide light curves of their interest as a starting point; then our model starts gathering the stars that are most likely to belong to the same cluster.

During the search of candidates, having a suitable light curve similarity function is mandatory. Light curves are highly heterogeneous; they present a different number of observations and different lengths in time domain among exemplars. Light curves are also unevenly sampled and often present gaps, which makes difficult to apply directly general-purpose and traditional time series techniques (Mackenzie et al. 2016; Richards et al. 2011; Huijse et al. 2014). Most similarity functions coming from the machine learning community are not designed to work with variable stars. Light curves have to be transformed into vector-like data before being passed to a similarity function. Traditionally, to do this, a series of descriptors called *features* are extracted from the original light curves, so that each gets represented by a feature vector (Nun et al. 2015). Features vary in nature, and there has been extensive research on light curve feature extraction (Debosscher et al. 2007; Kim et al. 2009, 2011; Pichara et al. 2012; Kim et al. 2014; Richards et al. 2011; Huijse et al. 2012; Nun et al. 2015; Mackenzie et al. 2016). Although some of these methods have proven largely successful, they demand a lot of computing time and are mostly designed for supervised classification, where the construction of vector-like data can be separated from the classification process (Nun et al. 2015; Debosscher et al. 2007).

Our similarity function consists of organising the light curve data into a data structure we call *Variability Tree*. The *Variability Tree* allows us to create a new compressed representation of light curves, enough to capture the underlying structure that plays the most relevant role in the unsupervised classification process. This method produces a vector-like representation and integrates both ideas: feature-extraction and fast search, without having to tackle both problems separately. Although our method has been specially designed for light curves, taking into account their irregular nature, it can also be used on any time series dataset with even or unevenly sampled data and arbitrary length.

We present an experimental analysis with OGLE (Udalski et al. 1996, 2008), MACHO (Alcock et al. 1997, 2003) and KEPLER (Koch et al. 2010) light curve datasets, showing that the running time of the Variability Tree performs significantly faster than traditional feature extraction approaches, while retrieving more similar light curves.

The remainder of this paper is organised as follows: Section 2 gives an account of previous work on time series representation and similarity search. Section 3 introduces relevant background theory. In Section 4, the proposed method is described in detail. Section 5 describes the datasets that were used in experiments. Section 6 gives an account of the software stack we use for implementing the method. Section 7 presents experimental results. Finally, in section 8, the main conclusions of the work are presented.

## 2 RELATED WORK

As we mentioned before, our unsupervised approach does not rely on performing an initial clustering of the data, but on carry on a search of lightcurves that are most similar to a set of query stars provided by the astronomer. Time series similarity search is a problem that has been extensively studied (Agrawal et al. 1993, 1995; Yi et al. 1997; Chan & Fu 1999; Wu et al. 2000; Keogh et al. 2001; Popivanov & Miller 2002; Kontaki & Papadopoulos 2004; Sart et al. 2010; Rakthanmanon et al. 2013; Wang et al. 2013; Baydogan & Runger 2016). However, to the best of our knowledge, most of the published work assume evenly spaced time series, which does not match the light curves context. Agrawal et al. (1993) were the first using the Discrete Fourier Transform (DFT) for time series search based on similarity. Agrawal et al. (1995) proposed an indexing model for time series based on the Dynamic Time Warping (DTW) distance measure using an R*-Tree. Chan & Fu (1999) introduced for the first time the use of the Discrete Wavelet Transform (DWT) as a replacement for DFT. Keogh et al. (2001) presented Piecewise Alternative Approximation, an alternative to DFT and DWT. Popivanov & Miller (2002) introduced the use of smooth wavelets in DWT time series similarity search. Kontaki & Papadopoulos (2004) proposed an index for similarity search in time series with streaming data. Sart et al. (2010) used GPUs and FPGAs to accelerate subsequence search using the Dynamic Time Warping similarity measure. Rakthanmanon et al. (2013) proposed a series of optimizations to existing methods, which led them to perform queries on massive datasets of up to a trillion elements. Baydogan & Runger (2016) introduced a new representation for time series, based on an autoregressive kernel, along with a similarity measure for this representation.

All the methods listed above assume evenly sampled data, either in the transformations or in the similarity measures they use. Therefore, they cannot be directly applied to light curves. To do this, light curves would have to be either interpolated and re-sampled, or binned. This could lead in many cases to bad approximations, since light curves are often poorly sampled and have gaps with no information.

Several works present feature extraction methods on light curves, mostly based on statistics, auto-correlation patterns, periodogram analysis, and variability measurements (Debosscher et al. 2007; Kim et al. 2009, 2011; Richards et al. 2011; Huijse et al.

2012; Kim et al. 2014). Unfortunately, as mentioned previously, highly descriptive features are often computationally expensive.

Our work is inspired on what Nister & Stewenius (2006) proposed for image retrieval. It also relies heavily on the idea of clustering of subsequences proposed by Mackenzie et al. (2016) for light curve feature extraction. Specifically, we borrow from the latter the ideas of subsequence sampling and using the Time Warp Edit Distance (TWED) (Marteau 2009) as a distance metric for light curves.

## 3 BACKGROUND THEORY

To understand how our method works, it is vital to understand the main concepts of Cluster Analysis and to be familiarized with the generic tree data structure. In this section, we briefly explain these topics and how they are important in the context of the *Variability Tree*. Also, we review some essential topics of text retrieval that inspire this work.

### 3.1 Clustering and $k$-medoids

Clustering is the task of identifying groups of similar objects in a data set, such that elements in the same group are similar to each other, while elements in different groups are dissimilar (Gorunescu 2011). To find this groups, called clusters, we need a clustering algorithm and a distance measure. A clustering algorithm is a set of rules to find clusters, while a distance measure is a function that tells us how different are two elements of the data set. For example, if the elements in our data set are vectors in $\mathbb{R}^n$, a valid distance measure is the Euclidean distance.

The selection of a specific clustering algorithm affects not only the procedure how the clusters are found but also the final result. Besides that, different clustering algorithms receive different parameters. For example, the popular $k$-means clustering algorithm receives as input the number of clusters $k$ to be found and retrieves a set of $k$ centres together with an assignment to one cluster for each element of the dataset, such that each of the $k$ centres correspond to the centroids—or means—of the clusters and each element belongs to the cluster with the closest centre.

In the construction of the *Variability Tree*, we need to apply a clustering algorithm to a set of light curve subsequences. These subsequences are fragments taken from the light curves in the dataset. Subsequences are also light curves themselves, but they have observations in a smaller period of time. For this purpose, we need to choose a similarity a measure that allows us to compare light curves and a clustering algorithm that lets us use such a similarity measure. The *Vocabulary Tree* (Nister & Stewenius 2006) on which the *Variability Tree* is inspired uses $k$-means for clustering, which works only in Euclidean spaces in its original form. Unfortunately, due to their irregularity, differences in number of observations, and shifted time among light curves, Euclidean distance is not a suitable similarity measure for variable stars. This suggest to seek for another clustering algorithm.

The $k$-medoids (or Partition Around Medoids) (Kaufman & Rousseeuw 1987) clustering algorithm is a modified version of $k$-means, which allows using an arbitrary measure of similarity by choosing data exemplars as cluster centres, instead of setting the centre on the true mean of the cluster, as $k$-means does. We use $k$-medoids because it lets us use whatever similarity measure we think is best to compare light curves. Also, $k$-medoids has the advantage that lets us choose the number of clusters, giving us more control

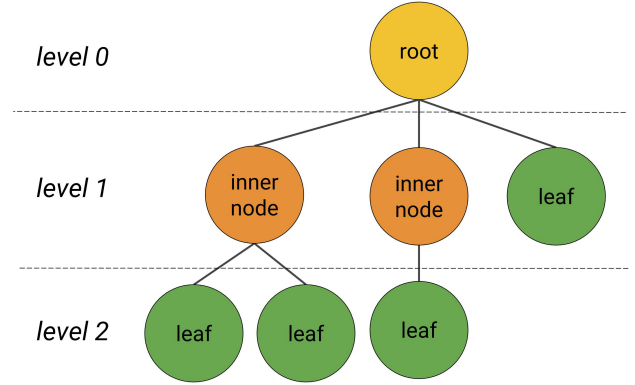

**Figure 1.** The generic tree data structure with its main parts.

over the structure of the tree, which doesn't hold for every clustering algorithm.

When using $k$-medoids, the pairwise distance between all the exemplars have to be calculated in advance and stored in a distance matrix. The distance matrix and the number of clusters $k$ are given to the algorithm as parameters. $k$ medoids identifies $k$ exemplars as centres and assigns every element to one of $k$ clusters.

### 3.2 Time Warp Edit Distance

We use the Time Warp Edit Distance (TWED) (Marteau 2009) as a distance measure for light curve subsequences. The TWED is a measure of distance for time series defined by the number of edit operations needed to transform one time series into another. It is designed to compare two time series, regardless of the number of data points each one has, which makes it perfect for irregular time series, such as light curves. It has already been successfully used by Mackenzie et al. (2016) for comparing light curves. A detailed explanation of TWED calculation can be found in Marteau (2009) and Mackenzie et al. (2016).

### 3.3 Trees

The *Variability Tree* is, as its name indicates, a tree data structure. In this section, we present the main characteristics of trees, as we understand them, together with relevant nomenclature that will be used in further sections.

A tree consists of a series of nodes which are connected by unidirectional links, following certain restrictions. Each node has a unique identifier and a value. The value can be any piece of data, depending on the purposes of the tree.

If node *A* has a link pointing to node *B*, we say *A* is the *parent* of *B* and *B* is *A*'s *child*. A node in the tree can have any number of children, but only one parent. Every node but one must have a parent. The node in the tree that doesn't have a parent is called the *root* of the tree and it is an *ancestor* of every other node in the tree. A node in the tree doesn't necessarily have to have children. If a node doesn't have any children, it is called a *leaf* of the tree. On the contrary, if it has children, and it is not the root of the tree, we call it an *inner node*.

We say a node belongs to *level n*, or has *depth n*, where *n* is the number of links between itself and the root node. In other words, the root node belongs to level 0, its children belong to level 1, and

**Table 1.** Text retrieval concepts exemplified.

| | |
|---|---|
| **D1** | A red car. |
| **D2** | A red apple and a red balloon. |
| **D3** | A yellow banana and a green apple. |

(a) Example of a documents dataset

| ID | word | IDF |
|---|---|---|
| 1 | a | 0.0 |
| 2 | and | 1.1 |
| 3 | apple | 0.4 |
| 4 | balloon | 0.5 |
| 5 | banana | 1.1 |
| 6 | car | 1.1 |
| 7 | green | 1.1 |
| 8 | red | 0.4 |
| 9 | yellow | 1.1 |

(b) A vocabulary for the documents dataset in a). Each word has an ID and IDF (calculated according to equation 1)

| | 1 | 2 | 3 | 4 | 5 | 6 | 7 | 8 | 9 |
|---|---|---|---|---|---|---|---|---|---|
| D1 | 1 | 0 | 0 | 0 | 0 | 1 | 0 | 1 | 0 |
| D2 | 1 | 0 | 1 | 0 | 1 | 0 | 0 | 2 | 0 |
| D3 | 2 | 1 | 1 | 1 | 0 | 0 | 1 | 0 | 1 |

(c) A BoW representation for the example dataset. Rows represent each document and columns represent each word (corresponding ID in the vocabulary). Each number counts the occurrences of the word in each document.

| | 1 | 2 | 3 | 4 | 5 | 6 | 7 | 8 | 9 |
|---|---|---|---|---|---|---|---|---|---|
| D1 | 0 | 0 | 0 | 0 | 0 | 1.1 | 0 | 0.4 | 0 |
| D2 | 0 | 0 | 0.4 | 0 | 0 | 0 | 0 | 0.8 | 0 |
| D3 | 0 | 1.1 | 0.4 | 0.4 | 1.1 | 0 | 1.1 | 0 | 1.1 |

(d) A TF-IDF representation for the dataset. Corresponds to the BoW representation where each number is multiplied by its respective IDF value.

so on. Figure 1 shows a visual representation of a tree, displaying its main parts and their relations.

Another concept worth mentioning is the *branching factor* of the node. This is the number of children a node has. Also, if every inner node of the tree has the same branching factor, it is common to say this number is the branching factor of the tree.

### 3.4 Text retrieval review

Although our direct inspiration comes from image retrieval, the main concepts of the Variability Tree have their origins on text retrieval. In this section, we will review key concepts of text mining and how they give origin to the Variability Tree.

When building a text retrieval framework, the first thing to be done is to create a numeric representation for the documents in the data set. There are several ways transform documents into numeric vectors, but, arguably, the most canonical model is Bag of Words (BoW). Typically, before building the representation, some preprocessing steps are performed to the text, but these are not relevant for our purposes.

To transform documents into vectors using BoW, first, a dictionary is built by assigning a unique identifier to each of the words that appear in the set of documents. Then, each document in the dataset is represented by a vector, that indicates how many times each word of the dictionary was mentioned in it. Usually, the identifier of the word serves as the index and the number of repetitions is stored in each component.

More complex representations use weighing schemes, that assign a value of importance to each word of the dictionary. One of the most widely used weighing schemes is *tf-idf* (Salton & McGill 1983). This scheme is based on the notion that an infrequent word gives more information than a frequent one. For each word, we calculate a weight called *inverse document frequency (idf)*. This weight depends on the inverse of the number of documents the word appears:

$$\text{idf}(word) = \ln \frac{N}{d(word)} \quad (1)$$

Where $N$ is the number of documents (3 in our example) and $d(word)$ corresponds to the number of documents in which *word* appears. See Table 1 for an example.

A main component of text retrieval engines are inverted indices. Inverted indices—also known as inverted files—work like a glossary of terms, where you can look up a word and be directed to the places where it was mentioned. An inverted index is a data structure that stores for each word in the dictionary, the list of documents where it was mentioned, together with the number of occurrences of the word in the documents. This makes possible to answer the question of which documents mention a certain word $w$, without having to search document by document if $w$ is present.

With documents represented as vectors and having built inverted files, we can perform text relevance queries. A query consists on a phrase and we expect the text retrieval engine to retrieve us a list of documents from the dataset sorted by relevance. To perform a query, first, the query is represented by a factor, using *BoW* or *tf-idf*, the same way as documents in the dataset. Then, inverted files are used to find which documents contain the words in the query phrase. Finally, documents are sorted according to their similarity to the query vector. Usually vectors are previously normalized, and the angle between them is used as a measure of similarity.

### 3.5 From the Vocabulary Tree to the Variability Tree

The approach above was shown to be useful not only for text retrieval. Sivic & Zisserman (2003) built an image retrieval framework using the same approach. In their work, the authors propose an image representation as a document composed of *visual words*. To do this, they extract features from images and build a dictionary of *visual words* using $k$-means to cluster the features. In this framework, each cluster represents a *visual word*. An occurrence of a *visual word* $w$ in a document $d$ is when a feature extracted from $d$ belongs to the cluster $w$. Based on this, they extrapolate to images the approach described above, using *tf-idf* for vector representation and inverted files for fast retrieval.

In Nister & Stewenius (2006), the authors present the *Vocabulary Tree*, an image recognition framework that uses the text retrieval approach that was brought to computer vision by Sivic & Zisserman (2003). They also build a dictionary of visual words, but instead of performing $k$-means once, they build a tree structure by dividing clusters into sub-clusters repeatedly, where each sub-cluster of any of the clustering levels represents a visual word. This results in a larger vocabulary and is shown to improve retrieval results.

The *Variabilty Tree* extends the *Vocabulary Tree* approach to light curves. In the following section we explain how the tree is built by clustering subsequences of light curves, allowing fast light curve retrieval.

# 4 METHOD DESCRIPTION

We can divide our method into two significant steps; the first one is the construction of a data structure (*Variability Tree*) that allows a fast light curve search. The second step is the algorithm that returns a list of candidate light curves. These light curves are the ones that most likely belong to the class of the query star, ranked by probability.

The *Variability Tree* is a data structure that allows performing fast similarity search and retrieval from light curve datasets. In this section, we describe in detail how the *Variability Tree* is built from the light curve data set and present an algorithm to perform queries over the tree.

It is important to note that the construction of the data structure is an *off-line* process. This means it is done in advance to the queries and does not have to be repeated each time a query is performed.

The construction of the *Variability Tree* consists of five steps. (i) First, we standardise the amplitude of the light curves in the dataset. (ii) In the second step, we take from the dataset a random sample of fragments of time series called subsequences. (iii) In the third step, these subsequences are hierarchically clustered with $k$-medoids, creating a tree structure. (iv) The fourth step consists in running a sliding window over every time series in the dataset and adding the resultant subsequences to the tree that has been previously built (see figure 4). (v) Finally, in the fifth step, we build a matrix representation for the dataset, based on the count of their subsequences in the tree nodes. These five steps are described in Sections 4.1.1, 4.1.2, 4.1.3, 4.1.4 and 4.1.5 respectively.

The algorithm to query the *Variability Tree* has five main steps (see figure 2). (i) The first step consists in standardising the query light curve and extracting subsequences from it, as it was done before with the light curve in the dataset. (ii) The second step consists in adding the extracted subsequences to the tree. (iii) In the third step, a vector representation of the query is built, based on the number of subsequences that reached each tree node. (iv) The fourth step consists in creating a reduced version of the query vector and the matrix representation of the dataset. This step is the key point of this method to achieve fast results. (v) Finally, in the fifth step, we calculate the distances between the reduced query vector and the rows of the query matrix—representing the light curve in the data set—, and these distances are sorted to retrieve the result in the form of a ranking.

These five steps are described in Sections 4.2.1, 4.2.2, 4.2.3, 4.2.4, 4.2.5.

## 4.1 Building the *Variability Tree*

To build the *Variability tree*, we start with a data set of light curves that are in the same band, but can have different lengths and represent objects of different nature.

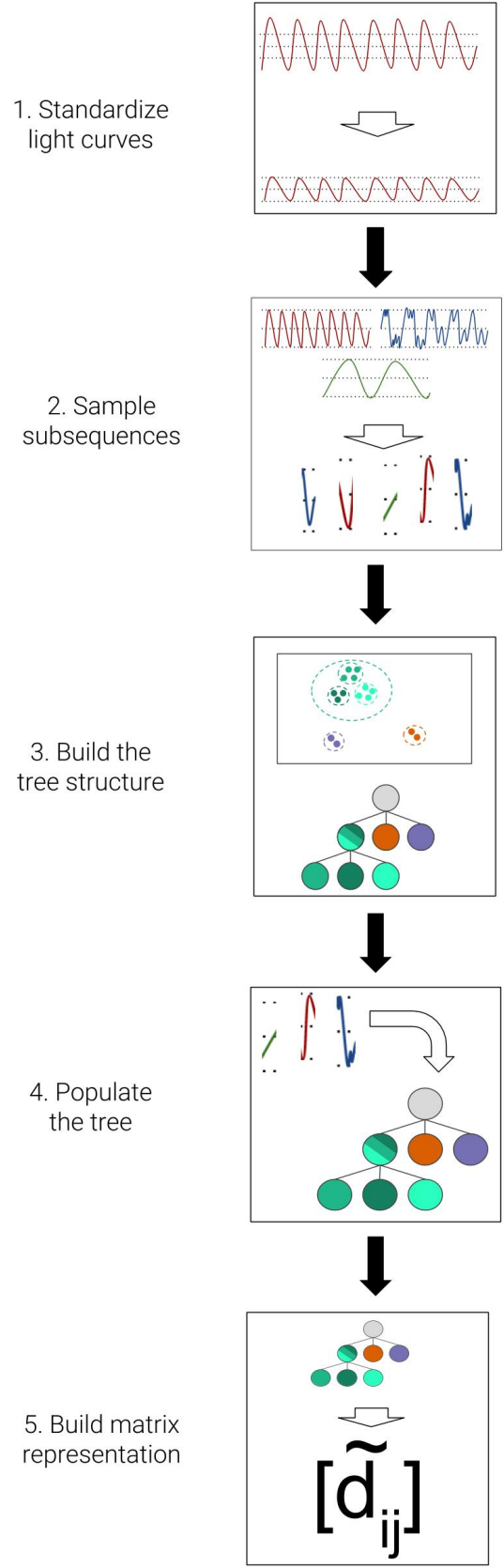

**Figure 2.** The *Variability Tree* building process.

*4.1.1 Standardise light curves*

The first step to build the *Variability Tree* is to standardise the magnitudes of all the light curves in the dataset. We subtract their mean magnitude and divide by their standard deviation, leaving every light curve with a mean magnitude of 0 and standard deviation of 1. Despite losing information about the absolute magnitude of the light curves, we found that better results were achieved with standardised light curves.

*4.1.2 Sample subsequences*

In this step, we take a random sample of $N$ light curves from the data set. From each of the sampled light curves, we extract a fragment from a time window of size $t_w$. The starting point of this time window is taken randomly. We call these fragments *subsequences*. $N$ and $t_w$ are parameters of the model, so their values have to be given by the user. $N$ must be lower than the total number of light curves in the data set.

Since the light curves are not uniformly sampled, the extracted subsequences do not necessarily have the same number of data points. The parameter $t_w$ defines the size of the window in terms of time and not in terms of the number of observations.

In subsequent steps, we will need to compare these subsequences using the Time Warp Edit Distance (TWED) measure (explained in Section 3.2). In order to do so, we shift the time values of the extracted subsequences, setting their initial values to zero.

If we do not have labels for the dataset, these $N$ light curves are uniformly sampled from the data set, leading to a completely unsupervised model. On the other hand, if the data set is labelled, we can optionally perform a stratified sampling. This means that we sample the light curves in a way that the size of the sample taken from each class is proportional to the total number of light curves in it. The goal here is to build a balanced sample, that is representative of the class distribution in the dataset.

*4.1.3 Build the tree structure*

In this step, we use the $k$-medoids clustering algorithm to build a tree data structure from the previously obtained subsequences. To do this, we calculate the pairwise TWED distance between the extracted subsequences and store them in a matrix.

After doing this, we run $k$-medoids to cluster the subsequences. The algorithm receives the distance matrix and the number of clusters $k$ as input.

As a result, this algorithm identifies $k$ of the $N$ subsequences as cluster centres and each of the $N$ subsequences as part of one cluster. The clustering algorithm is run again over each of the obtained clusters, creating subclusters. The process of subdividing clusters into smaller clusters is repeated $m$ times, dividing each subcluster into $k$ smaller subclusters. This is applied to every subcluster unless it has less than $k$ elements and cannot be subdivided.

This resulting hierarchy of $m$ levels of clusters and subclusters serves us to build the structure of the *Variability Tree*. In this tree, each of the clusters and subclusters is represented by a node. When a cluster is divided into subclusters, these are represented by the children of its respective node.

In each node of the tree, we store the centre of its respective cluster—that is, the subsequence which is at its centre—. Figure 3 illustrates how a tree is built from a hierarchy of clusters.

Both $k$ and $m$ are parameters of the *Variability Tree* and, therefore, have to be defined by the user. These parameters determine the number of nodes it will have. The parameter $k$, which is the number of clusters in $k$-medoids, becomes the *branching factor* of the tree; while $m$ defines *maximum depth* of the tree. A tree with branching factor $k$ and maximum depth $m$ will have a maximum number of $\sum_{i=0}^{m} k^i$ nodes.

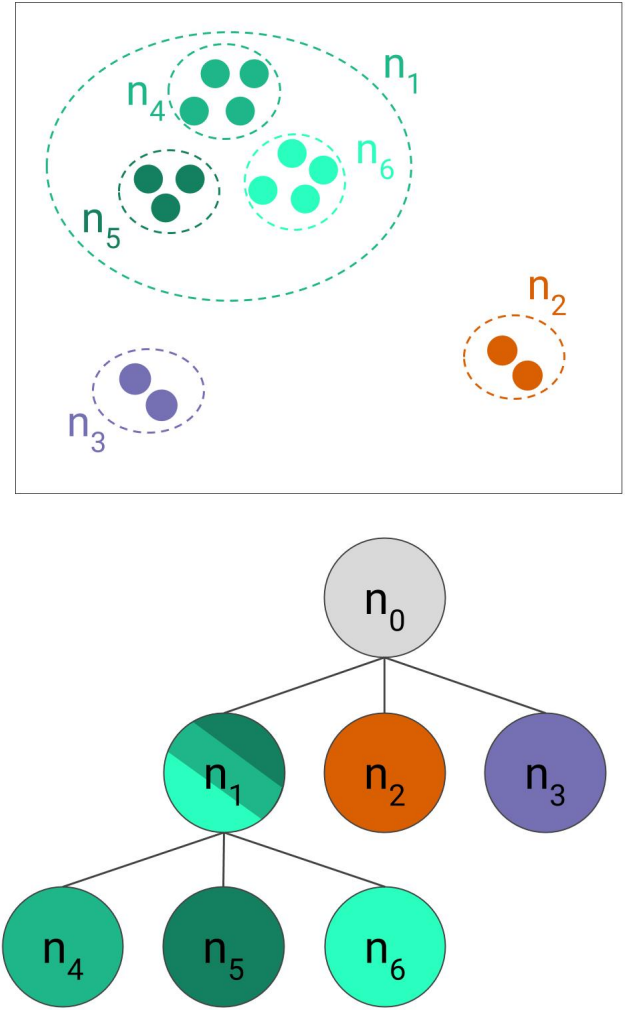

**Figure 3.** Tree construction through top-down hierarchical $k$-medoids. For simplicity, light curves are represented as points in a 2-dimensional space. In this example, branching factor $k$ is set to 3 and maximum level $m$ is set to 2, leading to a maximum possible of $\sum_{i=0}^{2} 3^i = 13$ nodes. Since clusters at nodes $n_2$ and $n_3$ have less than $k$ elements, they do not split into subclusters, resulting a total of 7 nodes. Clusters at nodes $n_4$, $n_5$ and $n_6$ do not split because they belong to level 2, which is set as maximum.

The idea behind clustering subsequences is to find subsequences that could represent building blocks for light curves, the same as words are the building blocks of documents (see Section 3.4). This process is analogous to what is done in image retrieval by clustering previously extracted features to find the best *visual words* to describe them.

*4.1.4 Populate the tree*

Up to this point, we have built the structure of the *Variability Tree*. Now it has to be populated with all the light curves in the data set. This is done by extracting a series of subsequences from each of the light curves in the data set and adding these subsequences to the tree.

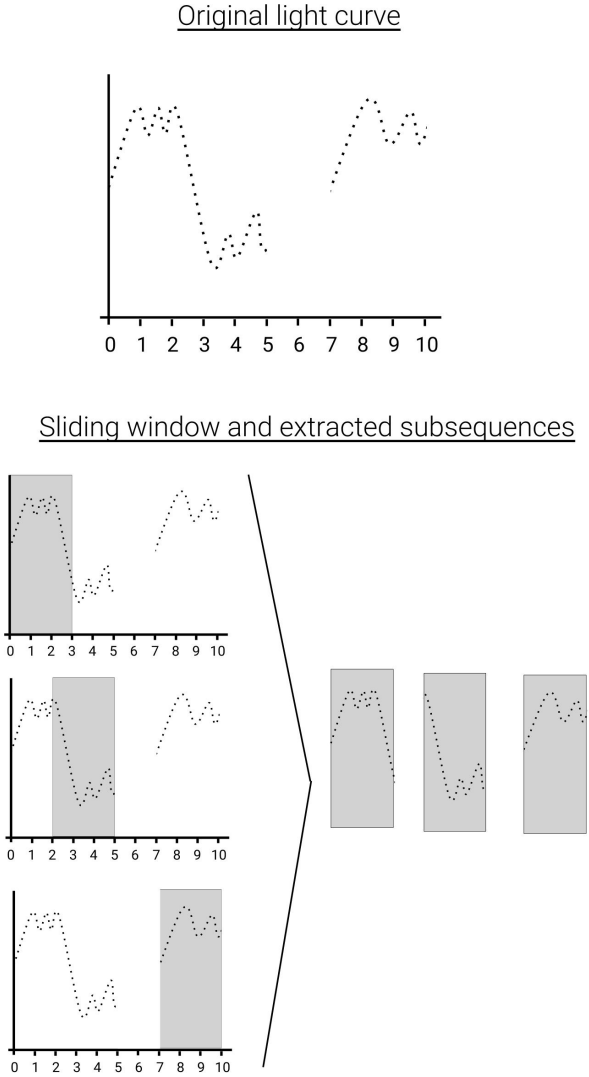

**Figure 4.** Subsequences are extracted from light curves using a sliding window.

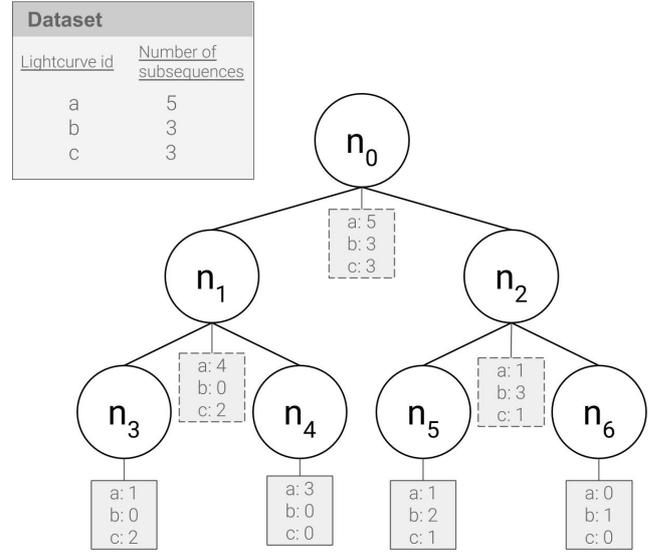

**Figure 5.** Variability Tree after being populated with full dataset. The number of subsequences taken from each light curve is not necessarily the same and depends on the length of the light curve. Each leaf node ($n_3$, $n_4$, $n_5$ and $n_6$) stores a register of how many subsequences of each lightcurves reached there. For example, $n_3$ was reached by 1 subsequence of light curve **a**, 2 of light curve **c** and none of light curve **b**. Non-leaf nodes have virtual dictionaries that are calculated by adding up the values of their children's dictionaries.

To extract subsequences from a light curve, we run a sliding window of time-step $t_s$ and time-window $t_w$ across it. This sliding window starts at the beginning of the light curve, taking a subsequence that contains all data points between its first observation, at $t_0$ and $t_0 + t_w$. Then, it slides across the light curve and extracts a subsequence between the time of first observation after $t_s$, let's say $t'_s$, and $t'_s + t_w$. The window keeps sliding by a time-step of size $t_s$ and extracting subsequences until it reaches the end of the light curve. This means the number of subsequences extracted for each light curve depends on the length of the light curve.

The parameter $t_w$ is the same that was mentioned in Section 4.1.2. The value of $t_s$ is also a parameter of the model; and, as well as $t_w$, is fixed in time and not in number of observations. Usually, $t_s$ should be much smaller than $t_w$ so that the sliding window produces overlapping subsequences. This is done to avoid missing any patterns in the light curve.

Subsequences are added into the tree through its root and they are propagated down through the node representing their closest cluster centre until they reach a leaf. This is done in the following manner: first, we calculate the *TWED* distances between the subsequence we want to add and each of the subsequences stored in the first level of the tree— i. e. the centres of the clusters found by $k$-medoids on its first execution—. After this, we select the node whose centre has the smallest distance to the subsequence. If this node is not a leaf, we repeat the process and calculate the distances between the subsequence and the subsequences stored in the children of the selected node. We repeat this process consecutively until the added subsequence reaches a leaf.

In each leaf node of the tree, we keep a record of how many subsequences of each light curve have reached that leaf. Initially, all these records are empty. Also, non-leaf nodes have records that show how many subsequences of each light curve have passed through them. Records in non-leaf nodes are virtual and not stored in memory because they can be easily obtained by adding up the records of their children nodes. This makes the *Variability Tree* use much less memory than it would if all registers were physically stored. Figure 5 shows how a *Variability Tree* looks after being populated with a dataset of three light curves.

These records work as inverted indices. As explained in Section 3.4, inverted indices allow fast retrieval by storing for each of the words in the dictionary, the list of documents that mention that word and the number of occurrences. What we are doing with this records is exactly the same if we consider that the subsequences stored in the nodes of the tree represent the building blocks of light curves, the same as words are the building blocks of documents.

*4.1.5 Build matrix representation*

Once the *Variability Tree* has been built and populated, we store the information gathered in the data structure in a matrix, which will help us on the querying stage that will be explained further on.

We build a matrix $D$ of $l \times n$ dimensions, where $l$ is the number of light curves in the original database and $n$ is the number of nodes in the tree. In the position $D_{ij}$ of the matrix, we store the number of light curve subsequences $i$ that passed through node $j$. In other words, each light curve in the original dataset is represented by a row in matrix $D$, indicating how many of its subsequences passed through each of the nodes of the *Vocabulary Tree*. This is a sparse matrix, meaning that many of its positions have zeroes. When implementing the method, a sparse matrix representation has to be used, since traditional matrix representations will use unnecessary amounts of space to store null positions.

In this step, we also calculate a weight for each node in the tree. As explained in 3.4, the *tf-idf* weighing scheme assigns more weight to a word that appears in fewer documents. Applying the same principle, we assign more weight to a node where subsequences of fewer light curves have passed and assign less weight to crowded nodes.

The weight of node $j$ is calculated by:

$$w_j = \ln \frac{l}{l_j} \quad (2)$$

where $l$ is the total number of light curves in the data set and $l_j$ is the number of light curves with at least one subsequence that passed through node $j$. We define $w_j = 0$ when $l_j = 0$.

We use these weights to obtain a weighted version of the matrix $D$. We denote it as $\bar{D}$ and its elements are defined as follows:

$$\bar{d}_{ij} = w_j d_{ij} \quad (3)$$

Finally, the rows in matrix $\bar{D}$ are divided by its norm. As will be explained later, having rows as unitary vectors will allow us to perform fast distance computations. Also, we don't lose relevant information by doing this, since we are not interested in the absolute count of subsequences in a node, but the relative count—that is, the number of subsequences that passed through one node compared to the number of subsequences that passed through the others—.

**4.2 Light curve retrieval: querying the tree**

As stated in Section 1, the input for a query can be any light curve—either one that is already contained in the dataset or not—and its output is a list of light curves from the data set ranked by its similarity to the input. We now explain the five steps that are performed by the model, starting from a single target light curve and ending in a ranking (see figure 6).

*4.2.1 Standardise query light curve and extract subsequences*

The first thing we do with the target light curve is to standardise its amplitude by dividing it by its mean magnitude and dividing by its standard deviation. After that, we extract subsequences from the target light curve using a sliding window of time-step $t_s$ and time-window $t_s$. These two procedures are identical to what was done with the data set light curves and explained respectively in Sections 4.1.1 and 4.1.4.

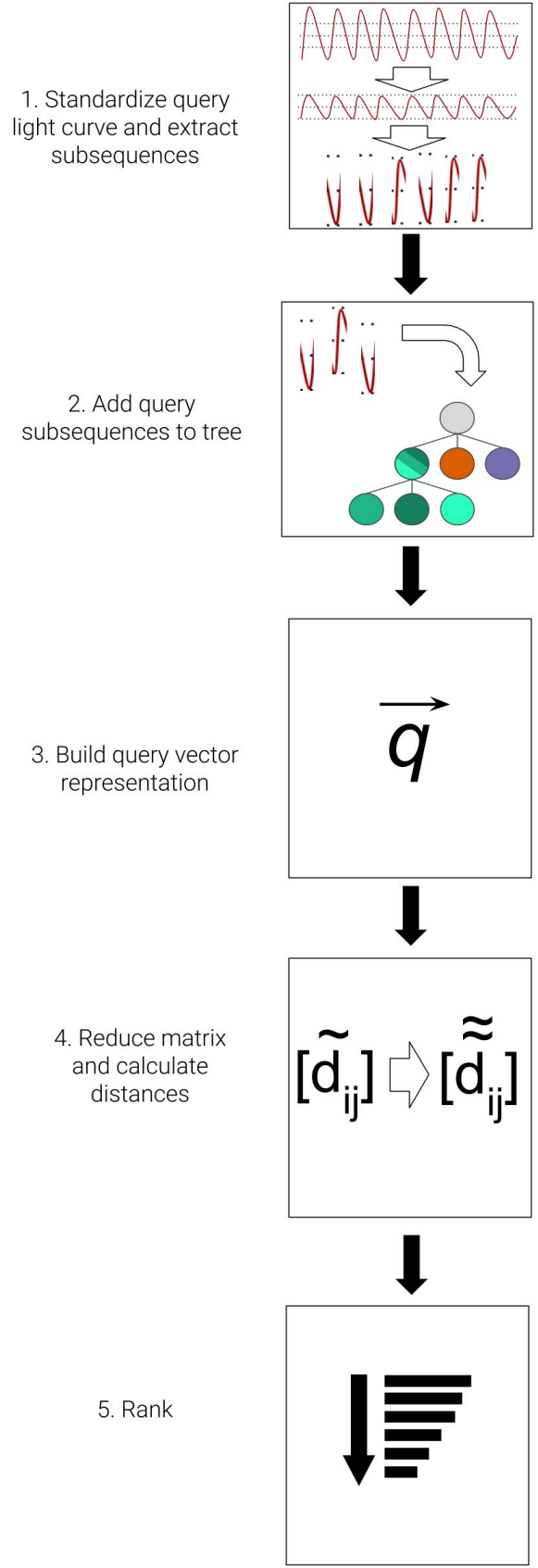

**Figure 6.** The *Variability Tree* querying process.

*4.2.2 Add query subsequences to tree*

What we do in this step is also very similar to what was done previously to the light curves in the data set.

The subsequences that we have just extracted are added to the tree until each reaches a leaf, the same way it was previously done with the light curves in the database, as explained in Section 4.1.4. In each node, we keep a record of how many subsequences of the query light curves passed through it. This record is kept separately from the one that stores the number of subsequences from the data set light curves.

*4.2.3 Build query vector representation*

After adding to the tree the subsequences extracted from the query light curve, we copy the query data that we recorded in the tree to a vector **q**.

This vector stores how many subsequences of the query light curve passed through each node of the tree, and it is weighted in the same manner as rows in matrix $\bar{D}$. Its components are defined by the following relation:

$$q_j = w_j p_j \quad (4)$$

where $w_j$ is the weight of node $j$ and $p_j$ is the number of subsequences of the query light curve that passed through node $j$. After this, the query vector **q** is normalized, as it was done with the rows of the matrix $\bar{D}$.

*4.2.4 Reduce matrix and calculate similarity*

Now that we have the data set light curves represented as the rows of matrix $\bar{D}$ and the query light curve represented as vector **q**, we can calculate a similarity measure between them. To do so, we use the cosine similarity, which is used in the text retrieval approaches we are being inspired by (Sivic & Zisserman 2003; Nister & Stewenius 2006). The cosine similarity between two vectors is defined as the cosine of the angle between them. That is,

$$\text{similarity}(\mathbf{a}, \mathbf{b}) = \frac{\mathbf{a} \cdot \mathbf{b}}{\|\mathbf{a}\| \, \|\mathbf{b}\|} = \cos(\theta) \quad (5)$$

where $\theta$ is the angle between **b** and **q**.

Since the rows of matrix $\bar{D}$ and the query vector **q** are already normalized, the calculation of the cosine similarity is reduced to the calculation of the dot product between them. In other words, $\bar{D}\mathbf{q}$ will result in a vector of length $l$, with the similarity between the query vector and each of the light curves in the data set.

Although this operation gives us effectively a similarity measure between the query and every light curve in the data set, it can be improved by avoiding to do some unnecessary calculations. Usually, there will be several light curves in the data set that do not share any node in the tree with the query. In these cases, the dot product between **q** and its respective row will be zero. Since we only want to get similar light curves, these light curves are not relevant in our resulting ranking. Using the inverted indices we kept in the leaves of the trees, we can select from matrix $\bar{D}$ only the rows that correspond to light curves that share at least one node with the query. We can build a new matrix selecting only the relevant rows.

Our $\bar{D}$ matrix has also columns that can be ignored to calculate the dot product. Usually, there will be nodes where none of the subsequences from the query light curve passed. The columns in $\bar{D}$ representing these nodes can be ignored since their values will be multiplied by zeroes when calculating their cosine similarity to **q**.

Table 2. Class distribution of OGLE labelled set.

| Class Name | Abbreviation | No. of elements |
|---|---|---|
| Classical Cepheids | CEP | 8026 |
| Type II Cepheids | T2CEP | 609 |
| Anomalous Cepheids | ACEP | 83 |
| RRLyrae | RRLYR | 44262 |
| Long Period Variables | LPV | 343816 |
| Double Period Variables | DPV | 137 |
| Delta Scuti | DSCT | 58 |
| Eclipsing binaries | ECL | 43859 |

Taking this into consideration, we build a reduced matrix $\bar{\bar{D}}$ by selecting from $\bar{D}$ only its relevant rows and columns. This can be done by a fast matrix slicing operation. We also build a reduced query vector $\bar{q}$, by eliminating its null positions—that is, the positions representing nodes where no query subsequences passed—.

Now that we have removed unnecessary data, we calculate $\bar{\bar{D}}\mathbf{q}$ to obtain the similarity between the query and every light curve in the dataset that has a cosine similarity larger than zero.

*4.2.5 Rank*

Finally, the ids of the light curves in the data set are sorted according to their similarity to the query. These sorted ids, together with their similarity scores constitute the output of the querying algorithm.

## 5 DATA

For experimental purposes, we used photometric data from OGLE, MACHO and KEPLER surveys. In Section 5.1, we present the nature and composition of these datasets, while in Section 5.2 we present the composition of the datasets we used to query the model.

### 5.1 Datasets

*5.1.1 Ogle III Catalog of Variable Starts (OGLE)*

The OGLE Experiment (Udalski et al. 1996) is an ongoing project that started on 1992 aiming to find microlensing events on the Galactic Bulge and the Magellanic Clouds. Its third phase, OGLE-III, surveyed the sky between 2001 and 2009, producing a large amount of labelled photometric data. These data were published as the Ogle III Catalog of Variable Starts (OIII-CVS) (Udalski et al. 2008).

The published light curves are in two bands: I-band and V-band. However, we only used the I-band for our experiments, since light curves in the V-band did not have enough observations for our purposes.

For our experiments, we used 444,503 light curves from OIII-CSV, from which 440,850 were labelled. The class distribution of the labelled subset is shown in Table 2.

*5.1.2 The MACHO project*

The MACHO project (Alcock et al. 1997) started on 1992 with the objective of finding massive compact halo objects that would explain the presence of dark matter in the Milky Way. As a by-product, it produced more than 80 billion photometric measurements from the Galactic Bulge, the Large Magellanic Cloud and the Small Magellanic Cloud (Alcock et al. 1999). Light curves were obtained on two bands: blue and red. On 2001, the MACHO Collaboration Team published a catalogue that included labels for 21,474 variable stars in the Magellanic Clouds (Alcock et al. 2003).

**Table 3.** Class distribution of MACHO labelled set.

| Class Name | Abbreviation | No. of elements |
| --- | --- | --- |
| RRLyrae ab | RRL AB | 7405 |
| RRLyrae c | RRL C | 1765 |
| RRLyrae e | RRL E | 315 |
| Cepheid Fundamental | Ceph Fund. | 1185 |
| Cepheid 1st Overtone | Ceph 1st. | 683 |
| Long Period Variable WoodA | LPV A | 315 |
| Long Period Variable WoodB | LPV B | 822 |
| Long Period Variable WoodC | LPV C | 1134 |
| Long Period Variable WoodD | LPV D | 778 |
| Eclipsing Binary | EB | 6835 |
| RR Lyrae and GB blends | RRL + GB | 237 |

For our experiments, we used a set of 19,927,044 light curves from MACHO survey, corresponding to objects on the Large Magellanic Cloud, from which 20,391 were labelled.

The class distribution of the labelled subset is shown in Table 3. It is worth noting that in this dataset RRLyrae stars are subdivided into three subcategories and Long Period Variable are subdivided into four subcategories, while these are considered two big categories in our OGLE dataset.

#### 5.1.3 The Kepler Mission

The Kepler mission (Koch et al. 2010) was launched on 2009 by NASA with the main purpose of finding Earth-size exoplanets. As a result, it produced a huge amount of high-resolution photometric measurements.

Kepler light curves were published in two cadences: Short Cadence (SC), which has observations every 58.89 seconds, and, and Long Cadence (LC), which has observations every 29.4 minutes. It is worth noting that both LC and SC data have a much shorter cadence than OGLE and MACHO data. While MACHO has approximately one observation per 2 days and OGLE has around one observation per 6 days, Kepler long cadence data has around 50 observations per day. To make Kepler data more manageable, we downsampled the light curves to get one observation per day. We performed our experiments with these downsampled light curves. We used 207,617 light curves from this dataset. We didn't use any labelled data from this survey.

### 5.2 Query-sets

For each of the datasets mentioned in Section 5.1, we built a query-set. A query-set consists of a set of light curves for which we would perform unsupervised classification by obtaining their most similar light curves. We build the query-sets by taking a random sample of light curves for each of the classes in each dataset. Table 4 shows the composition of query-sets for each of the three dataset we use.

**Table 4.** Class distribution of query-sets built for OGLE, MACHO and KEPLER datasets.

| Class | Count |
| --- | --- |
| CEP | 100 |
| T2CEP | 98 |
| ACEP | 83 |
| RRLYR | 100 |
| LPV | 100 |
| DPV | 100 |
| DSCT | 51 |
| ECL | 86 |

(a) OGLE

| Class | Count |
| --- | --- |
| RRL AB | 100 |
| RRL C | 100 |
| RRL E | 100 |
| Ceph Fund. | 100 |
| Ceph 1st. | 100 |
| LPV A | 100 |
| LPV B | 100 |
| LPV C | 100 |
| LPV D | 100 |
| EB | 100 |
| RRL + GB | 100 |

(b) MACHO

| Class | Count |
| --- | --- |
| *Unlabelled* | 448 |

(c) KEPLER

**Table 5.** Values of parameters used at experiments.

| Parameter | Meaning | Value |
| --- | --- | --- |
| $N$ | Number of subsequences at sampling stage | 20,000 |
| $t_w$ | Size of subsequences in days | 250 |
| $t_s$ | Step of sliding window in days | 10 |
| $k$ | Branching factor of the tree | 3 |
| $m$ | Maximum level of Variability Tree | 10 |

## 6 IMPLEMENTATION

### 6.1 Software stack

Our implementation of the method is written in Python 3.5 using numpy (Van Der Walt et al. 2011), and numba (Lam et al. 2015) for efficient numerical computation, pandas (McKinney 2010) for data manipulation and scipy (Oliphant 2007) for sparse matrix representation. For K-Medoids, we use Szalkai (2013) implementation and for TWED we use Mackenzie et al. (2016). Our code is open-source under the MIT License and available at *https://github.com/luvalenz/time-series-variability-tree*.

### 6.2 Parameters

As explained previously in Section 4, the *Variability Tree* model has five parameters: the number of subsequences at sampling stage $N$, the size of the subsequences to be extracted $t_w$, the step of the sliding window $t_s$, the branching factor of the tree $k$ and maximum level of the tree $m$. To execute our python implementation, the user has to pass the values of this parameters as arguments to a python script. The values we used for each of the parameters in our experiments are shown in Table 5.

For the parameters that define how subsequences are extracted—that is, $N$, $t_w$ and $t_s$—, we chose to use the same values as Mackenzie et al. (2016). Although the method presented on their paper serves a different purpose than ours, both rely on clustering subsequences of light curves to build a new representation of the data. This led us to assume that the values that had proved successful in the subsequence extraction stage, would also be suitable for our work.

For the branching factor $k$, a larger value leads to a tree with more nodes per level. Such a tree preserves more information, but has a longer building time and longer query times, because more distance calculations have to be done each time a subsequence is added. On our experiments, we set $k$ to 3, allowing us to preserve the maximum amount of information while maintaining reasonable query times [1]. With the same purpose of maintaining reasonable query times, we set the maximum level of the tree $m$ to 10, setting a hard limit to the number of nodes in the tree.

# 7 EXPERIMENTAL RESULTS

In this section, we present the experiments we conduct to test the how the *Variability Tree* performs on unsupervised classification. We carry out experiments in two different set-ups. The first set-up (section 7.1) aims at measuring the performance of unsupervised classification using the *Variability Tree*. Here we use the rankings retrieved by the *Variability Tree* to classify light curves from labelled datasets. To do this, we work under the assumption that the most similar light curves to a given target should belong to the same class. In other words, if a light curve appears in the first $k$ positions of a ranking, our model predicts it has the same label as the target. To determine how good is the model to classify the light curves in the ranking, we measure the precision for different values of $k$. It is worth noting that in this set-up the labels are only been used for testing, but are not involved in the learning process, therefore, we can still talk about unsupervised classification. Our model only predicts that a certain light curves should share the same class, but if we do not previously know the class of one of them, we cannot predict the class of the others. The second set-up (section 7.2) aims at testing how unsupervised classification with the *Variability Tree* can scale to larger datasets. Here, we test our method with the full datasets presented in Section 5 (containing no labels). Since we do not have labels for this data, we cannot measure classification performance, but we report obtained rankings, together with the query times.

## 7.1 Experimental set-up No. 1: measuring ranking quality

In order to test how our model performs on unsupervised classification, we build a *Variability Tree* for each of the labelled dataset presented in Section 5.1. The first tree is built from 20,391 labelled light curves from MACHO dataset, while the second is built from 440,850 labelled light curves from OGLE dataset.

After this process, we use the previously built trees to perform a series of similarity searches (or queries). We perform similarity queries for light curves in the query-sets described in Section 5.2. For each light curve in the OGLE query-set, we obtain a ranked list of their most similar exemplars in the OGLE data set using the OGLE *Variability Tree*. The same process is repeated for MACHO.

The quality of these rankings is measured using **Precision at $k$ (P@k)**, which calculates the ratio of relevant results in the top $k$

[1] We consider a reasonable query time less than a minute for about ten millions of light curves. We use a server with 64 cores and 200 GB of RAM.

**Table 6.** Precisions at $k$ for rankings obtained in experimental set-up # 1. Values show mean and standard deviation.

|  |  | MACHO | OGLE |
|---|---|---|---|
| P@1 | VT | 1.00 ± 0.00 | 1.00 ± 0.00 |
|  | FATS | 1.00 ± 0.00 | 1.00 ± 0.00 |
|  | CBFL | 1.00 ± 0.00 | 1.00 ± 0.00 |
| P@5 | VT | 0.57 ± 0.30 | 0.59 ± 0.36 |
|  | FATS | 0.46 ± 0.27 | 0.53 ± 0.35 |
|  | CBFL | 0.50 ± 0.28 | 0.61 ± 0.37 |
| P@10 | VT | 0.50 ± 0.31 | 0.52 ± 0.40 |
|  | FATS | 0.39 ± 0.28 | 0.46 ± 0.38 |
|  | CBFL | 0.42 ± 0.28 | 0.54 ± 0.41 |
| P@20 | VT | 0.46 ± 0.32 | 0.49 ± 0.41 |
|  | FATS | 0.34 ± 0.27 | 0.42 ± 0.39 |
|  | CBFL | 0.37 ± 0.27 | 0.50 ± 0.43 |

positions of a ranking, that is,

$$P@k = \frac{|\{\text{relevant documents in first } k \text{ positions}\}|}{k} \quad (6)$$

In this set-up, we compare our results to the rankings obtained when performing nearest neighbour searches over previously transformed data sets using feature extraction methods. Here, we perform feature extraction on OGLE and MACHO labelled data, transforming the light curves into feature vectors. After that, we generate similarity rankings for the light curves in the query-sets by obtaining the closest vectors using the euclidean distance. We performed these nearest neighbour searches using a KD-Tree (Bentley 1975), which is a data structure specially designed to do this queries in a fast manner.

We use two different feature-sets to perform nearest neighbour searches and compare them to our proposed method. The first one corresponds to the feature set offered by the FATS Python library (Nun et al. 2015). This is a collection of 64 features used throughout literature. The second corresponds to a set of automatically learned features presented in Mackenzie et al. (2016) (Clustering Based Feature Learning (CBFL)), which is based on clustering of subsequences.

In Table 6, we show the P@k score with $k = 1, 5, 10, 20$ for rankings obtained using the *Variability Tree* (VT), as well as using nearest neighbour search over the FATS and CBFL feature sets. Results show that our method retrieves better quality rankings than nearest neighbour search in the MACHO datasets while obtaining comparable results to CBFL nearest neighbour search in OGLE. It is worth noting that the only reason why P@1 = 1.0 in all cases is that the query sets correspond to subsets of the data. Being built this way, all the query light curves are contained in the dataset, so for every query, the first element of the ranking—the most similar— should be the query light curve itself. That is indeed what happens, and works as a proof of concept that the searches are performed as they should.

While the quality of the obtained rankings can be comparable, the *Variability Tree* clearly outperforms nearest neighbour search in querying time for both datasets (OGLE and MACHO) and both feature sets (FATS and CBFL). Query time for these experiments are shown in Table 7.

Another result worthy of mention is that differences in precision between methods are not even amongst classes. In Figures 8 and 7, the distributions by class of P@10 are shown for OGLE and MACHO, respectively. For example, we can note that in MACHO the *Variability Tree* notoriously outperforms the other methods for the class *Ceph 1st*, while performing similarly to the others for the classes *LPV B* and *LPV C*.

**Table 7.** Query time (in seconds) for experimental set-up # 1. Values show mean and standard deviation.

|      | MACHO          | OGLE           |
| ---- | -------------- | -------------- |
| VT   | 2.37 ± 1.13    | 2.53 ± 1.02    |
| FATS | 80.78 ± 30.46  | 74.72 ± 56.07  |
| CBFL | 9.47 ± 3.50    | 8.07 ± 7.79    |

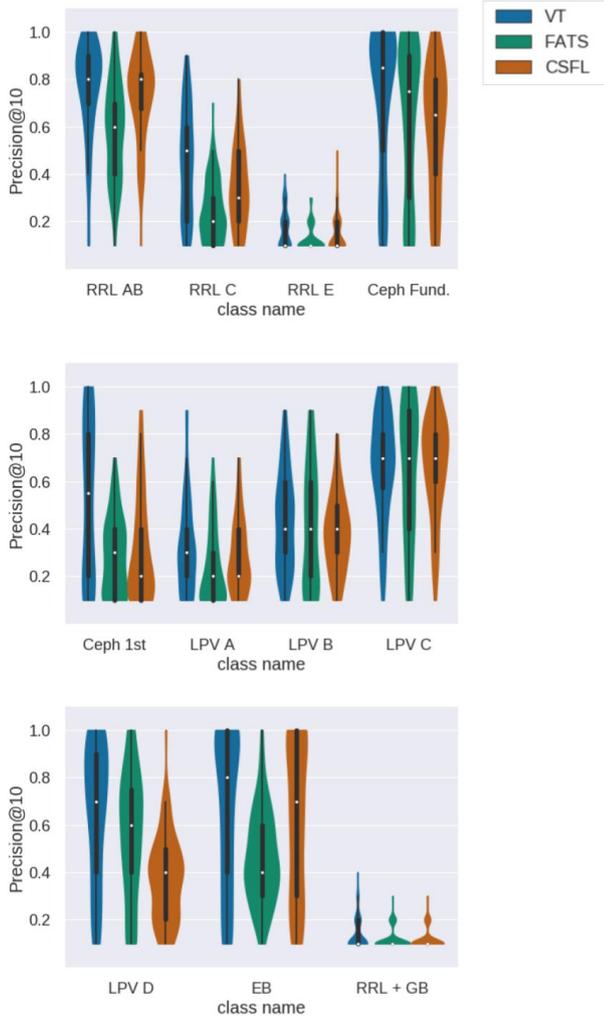

**Figure 7.** P@10 distribution by class for MACHO dataset.

We also find that light curves of some specific classes are likely to be found similar to light curves of other specific classes, while there are pairs of classes whose light curves are never found similar. To show this, we use our rankings to perform unsupervised classification. We classified the first five light curves of each ranking with the class of the target light light curve. The results of this classification are shown in Figure 9.

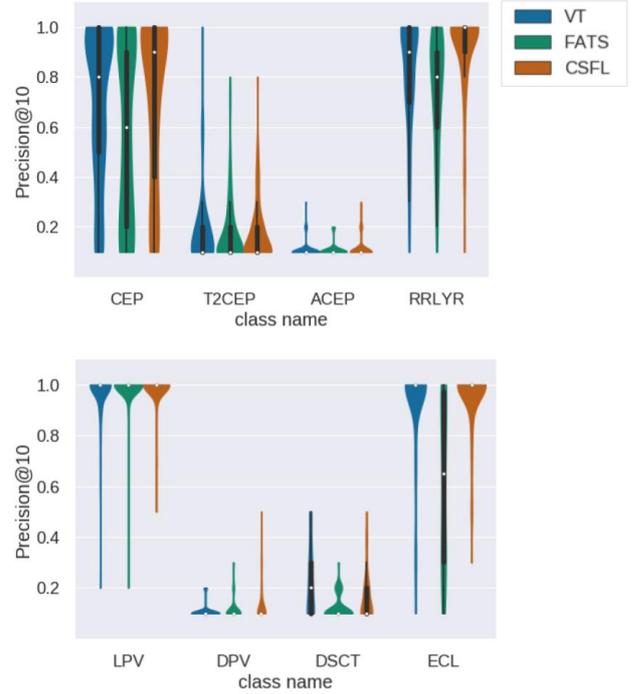

**Figure 8.** Precision@10 distribution by class for OGLE dataset.

**Table 8.** Average time in seconds for queries in experiments for unlabelled datasets.

| Dataset | Time [seconds]  |
| ------- | --------------- |
| MACHO   | 147.42 ± 48.42  |
| OGLE    | 3.62 ± 1.33     |
| KEPLER  | 2.95 ± 1.59     |

### 7.2 Experimental set-up No. 2: Large datasets

In these experiments, we use MACHO, OGLE and KEPLER datasets. For MACHO, we use 19,927,044 light curves, most of which were unlabelled. The few labelled light curves in the dataset correspond to the ones used in set-up No. 1. For OGLE, we use 444,503 light curves, for most of which we have labels. This dataset also correspond to a superset of the dataset used in 7.1. For KEPLER, we use an unlabelled dataset of 207,617 light curves.

In this set-up, we build a *Variability Tree* for each of these three datasets, using all the light curves mentioned in the above paragraph. After that, we query each tree with its corresponding query-set. For MACHO and OGLE, we use the same query-sets as in set-up No 1. For KEPLER, we use the query-set of 448 unlabelled light curves.

For these experiments, we calculate the query times, which are shown in Table 8, and store the obtained rankings. Figure 10 shows some examples of the rankings obtained in this set-up. Also, in https://github.com/<authorusername>/time-series-variability-tree , we make available for download the rankings for all of the queries performed in this experiment set-up, hoping they can be useful for the scientific community. Due to the lack of labels in this case, we cannot give a quality measure of the retrieved rankings, as we did in Section 7.1.

**Figure 9.** Confusion matrix for OGLE and MACHO.

## 8 CONCLUSIONS

In this work, we presented a novel data structure—the *Variability Tree*—tackles together the problems of feature-extraction and fast similarity search. We also showed how similarity search can be used to classify light curves without exhaustively training a specialized classifier.

Our results show that our method performs as well or better than nearest neighbour search over feature-vector data, while having a significantly lower computational cost.

We hope this work can aid astronomers to find objects of interest in existing datasets, that due to their size cannot be explored manually. We strongly believe that unsupervised methods, such as this, can help the task of classification, by separating interesting from uninteresting data.

Our Python implementation is open-source and available for download. The values of the parameters shown in Table 5 should be adjusted depending on the data.

**ACKNOWLEDGEMENTS**

We acknowledge the support from CONICYT-Chile, through the FONDECYT project number 11140643. This paper utilizes public domain data obtained by the MACHO Project, jointly funded by the US Department of Energy through the University of California, Lawrence Livermore National Laboratory under contract No. W-7405-Eng-48, by the National Science Foundation through the Center for Particle Astrophysics of the University of California under cooperative agreement AST-8809616, and by the Mount Stromlo and Siding Spring Observatory, part of the Australian National University.

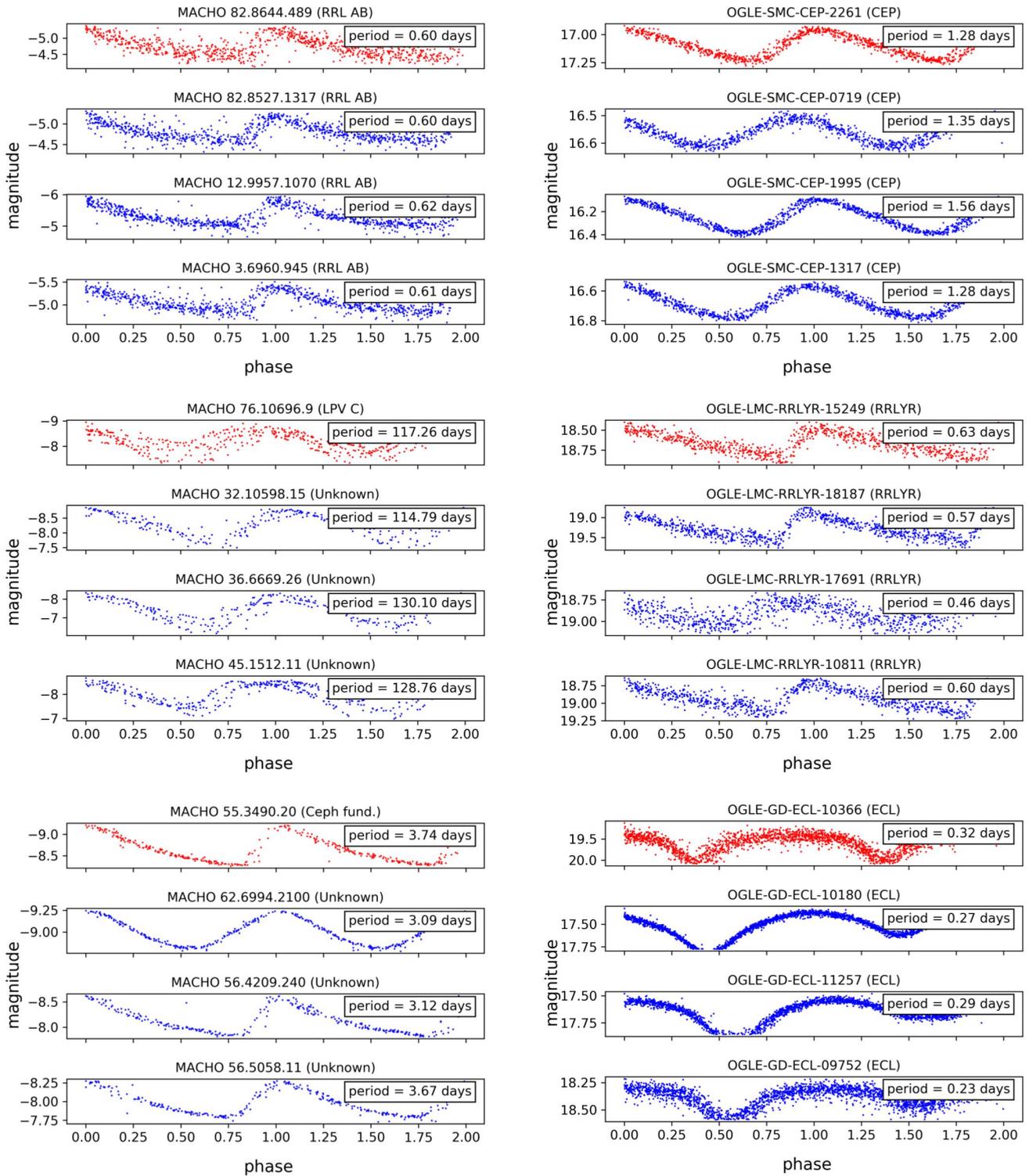

**Figure 10.** Sample queries on MACHO and OGLE data sets. The light curves are shown folded for visualization purposes, although we worked with unfolded raw data.

# REFERENCES

A. M., 1999, in Proceedings of the 11th Conference on Advances in Neural Information Processing Systems, NIPS. pp 543–549

Agrawal R., Faloutsos C., Swami A., 1993, Proceedings of the 4th International Conference on Foundations of Data Organization and Algorithms (FODO '93), 8958546, 69

Agrawal R., Lin K., Sawhney H. S., Shim K., 1995. pp 490–501, http://citeseerx.ist.psu.edu/viewdoc/download?doi=10.1.1.40.4034{&}rep=rep1{&}type=pdf

Alcock C., et al., 1997, The Astrophysical Journal, 486, 697

Alcock C., et al., 1999, Publications of the Astronomical Society of the Pacific, 111, 47

Alcock C., et al., 2003, Variable Stars in the Large Magellanic Clouds (MACHO, 2001). Vol. 2247, Washington, Smithsonian Institution Press.

Angeloni R., et al., 2014, Astronomy and Astrophysics, 567, A100

Baydogan M. G., Runger G., 2016, Data Mining and Knowledge Discovery, 30, 476

Bentley J. L., 1975, Communications of the ACM, 18, 509

Bhamidipaty A., Janakiraman A., Sarawagi S., Haritsa J., 2002, Workshop on Data Mining Lessons Learnt held in conjunction with the International Conference on Machine Learning

Bloom J. S., et al., 2012, Publications of the Astronomical Society of the Pacific, 124, 1175

Bradley P., Mangasarian O., 1998, in Proceedings of the International Conference on Machine Learning. Morgan Kaufmann, pp 82–90

Cabrera-Vives G., Reyes I., Förster F., Estévez P. A., Maureira J.-C., 2017, The Astrophysical Journal, 836, 97

Chan K.-P. C. K.-P., Fu A. W.-C. F. A. W.-C., 1999, Proceedings 15th International Conference on Data Engineering (Cat. No.99CB36337), pp 0–7

Debosscher J., Sarro L. M., Aerts C., Cuypers J., Vandenbussche B., Garrido R., Solano E., 2007, Astronomy and Astrophysics, 475, 1159

Elorrieta F., et al., 2016, Astronomy & Astrophysics, 595, A82

Förster F., et al., 2016, The Astrophysical Journal, 45, 1

Gorunescu F., 2011, Data Mining: Concepts and Techniques. Vol. 12, Springer (arXiv:1011.1669v3), doi:10.1007/978-3-642-19721-5, http://link.springer.com/10.1007/978-3-642-19721-5

Huijse P., Estevez P. A., Protopapas P., Zegers P., Pra??cipe J. C., 2012, IEEE Transactions on Signal Processing, 60, 5135

Huijse P., Estevez P. A., Protopapas P., Principe J. C., Zegers P., 2014, IEEE Computational Intelligence Magazine, 9, 27

Ivezic Z., et al., 2008, The Astrophysical Journal, p. 39

Kaufman L., Rousseeuw P., 1987, Clustering by means of medoids. Elsevier

Keogh E., Chakrabarti K., Pazzani M., Mehrotra S., 2001, Knowl. Inf. Syst., 3, 263

Kim D. W., Protopapas P., Alcock C., Byun Y. I., Bianco F. B., 2009, Monthly Notices of the Royal Astronomical Society, 397, 558

Kim D.-W., et al., 2011, The Astrophysical Journal, 735, 68

Kim D.-W., Protopapas P., Bailer-Jones C. A. L., Byun Y.-I., Chang S.-W., Marquette J.-B., Shin M.-S., 2014, Astronomy & Astrophysics, 566, A43

Koch D. G., et al., 2010, The Astrophysical Journal Letters, 713, 79

Kontaki M., Papadopoulos A., 2004. pp 63–72, doi:10.1109/SSDM.2004.1311194

Lam S. K., Pitrou A., Seibert S., 2015, Proceedings of the Second Workshop on the LLVM Compiler Infrastructure in HPC - LLVM '15, pp 1–6

Mackenzie C., Pichara K., Protopapas P., 2016, The Astrophysical Journal, 820, 138

Marteau P. F., 2009, IEEE Transactions on Pattern Analysis and Machine Intelligence, 31, 306

McKinney W., 2010, in van der Walt S., Millman J., eds, Vol. 1697900, Proceedings of the 9th Python in Science Conference. pp 51–56, http://conference.scipy.org/proceedings/scipy2010/mckinney.html

Nister D., Stewenius H., 2006, in Proceedings of the IEEE Computer Society Conference on Computer Vision and Pattern Recognition. IEEE, pp 2161–2168, doi:10.1109/CVPR.2006.264, http://ieeexplore.ieee.org/document/1641018/

Nun I., Pichara K., Protopapas P., Kim D.-W., 2014, The Astrophysical Journal, 79323

Nun I., Protopapas P., Sim B., Zhu M., Dave R., Castro N., Pichara K., 2015, pp 1–13

Oliphant T. E., 2007, Python for scientific computing. Vol. 9, doi:10.1109/MCSE.2007.58, , http://www.scipy.org/

Pichara K., Protopapas P., 2013, The Astrophysical Journal, 777, 83

Pichara K., Protopapas P., Kim D.-W. W., Marquette J.-B. B., Tisserand P., 2012, Monthly Notices of the Royal Astronomical Society, 427, 1284

Pichara K., Protopapas P., León D., 2016, The Astrophysical Journal, 819, 18

Popivanov I., Miller R., 2002, Proceedings 18th International Conference on Data Engineering, pp 212–221

Rakthanmanon T., Campana B., Mueen A., Batista G., Westover B., Zhu Q., Zakaria J., Keogh E., 2013, Transactions on Knowledge Discovery from Data TKDD, 7, 3047

Richards J. W., et al., 2011, The Astrophysical Journal, 733, 10

Salton G., McGill J. M., 1983, pp 24–51

Sart D., Mueen A., Najjar W., Keogh E., Niennattrakul V., 2010, in Proceedings - IEEE International Conference on Data Mining, ICDM. IEEE, pp 1001–1006, doi:10.1109/ICDM.2010.21, http://ieeexplore.ieee.org/document/5694075/

Sivic J., Zisserman A., 2003, Toward Category-Level Object Recognition, pp 1470–1477

Szalkai B., 2013, CoRR, abs/1304.6899

T. Z., R. R., M. L., 1996, in Proceedings of the ACM SIGMOD International Conference on Management of Data. pp 103–114

Udalski A., Szymański M., Kałuzny J., Kubiak M., Mateo M., 1996, Acta Astronomica, 42, 253

Udalski A., Szymański M. K., Soszyński I., Poleski R., 2008, Acta Astronomica, 58, 69

Van Der Walt S., Colbert S. C., Varoquaux G., 2011, Computing in Science and Engineering, 13, 22

Wang X., Mueen A., Ding H., Trajcevski G., Scheuermann P., Keogh E., 2013, Data Mining and Knowledge Discovery, 26, 275

Wu Y.-L., Agrawal D., El Abbadi A., 2000, in Proceedings of the 9th International Conference on Information and Knowledge Management (CIKM). ACM Press, New York, New York, USA, pp 488–495, doi:10.1145/354756.354857, http://portal.acm.org/citation.cfm?doid=354756.354857

Yi B.-K., Jagadish H. V., Faloutsos C., 1997, Data Engineering, pp 201–208

Z. H., 1998, Data Mining and Knowledge Discovery, 2, 283

If you want to present additional material which would interrupt the flow of the main paper, it can be placed in an Appendix which appears after the list of references.

This paper has been typeset from a TEX/LATEX file prepared by the author.